\documentclass[prl,twocolumn,aps,letterpaper, preprintnumbers,superscriptaddress]{revtex4}
\usepackage{hyperref}
\usepackage{verbatim}
\usepackage{amsmath}
\usepackage{latexsym}
\usepackage{revsymb}
\usepackage{yfonts}
\usepackage{ifthen}
\usepackage{tcolorbox}
\usepackage{natbib}
\usepackage{amsfonts}
\usepackage{amsmath}
\usepackage{amssymb}
\usepackage{amsthm}
\usepackage{graphicx}
\usepackage{bm}
\usepackage{bbm}
\usepackage{epsfig,color,amssymb}
\usepackage{subfigure}
\usepackage{amsfonts}
\usepackage{amscd}
\usepackage{amsmath}
\usepackage{multirow}
\usepackage{chemarrow}
\usepackage{dcolumn}
\usepackage{bm}
\usepackage{graphicx}
\usepackage{enumerate}
\usepackage{epsfig}
\usepackage{subfigure}
\usepackage{xcolor}
\usepackage{multirow}
\usepackage{ulem}
\usepackage{braket}
\usepackage{comment}
\usepackage{enumitem}
\usepackage{amsthm}
\usepackage{diagbox}

\renewcommand{\emph}[1]{\textit{#1}}

\begin{document}
\title{Topologically controlled multiskyrmions in photonic gradient-index lenses}

\author{Yijie Shen}\email{y.shen@soton.ac.uk}
\affiliation{Optoelectronics Research Centre, University of Southampton, Southampton SO17 1BJ, United Kingdom}
\author{Chao He}\email{chao.he@eng.ox.ac.uk}
\affiliation{Department of Engineering Science, University of Oxford, Parks Road, Oxford, OX1 3PJ, United Kingdom}
\author{Zipei Song}
\affiliation{Department of Engineering Science, University of Oxford, Parks Road, Oxford, OX1 3PJ, United Kingdom}
\author{Binguo Chen}
\affiliation{Guangdong Engineering Center of Polarization Imaging and Sensing Technology, Tsinghua Shenzhen International Graduate School, Tsinghua University, Shenzhen 518055, China}
\author{Honghui He}
\affiliation{Guangdong Engineering Center of Polarization Imaging and Sensing Technology, Tsinghua Shenzhen International Graduate School, Tsinghua University, Shenzhen 518055, China}
\author{Yifei Ma}
\affiliation{Department of Engineering Science, University of Oxford, Parks Road, Oxford, OX1 3PJ, United Kingdom}
\author{Julian A.J. Fells}
\affiliation{Department of Engineering Science, University of Oxford, Parks Road, Oxford, OX1 3PJ, United Kingdom}
\author{Steve J. Elston}
\affiliation{Department of Engineering Science, University of Oxford, Parks Road, Oxford, OX1 3PJ, United Kingdom} 
\author{Stephen M. Morris}
\affiliation{Department of Engineering Science, University of Oxford, Parks Road, Oxford, OX1 3PJ, United Kingdom}
\author{Martin J. Booth}
\affiliation{Department of Engineering Science, University of Oxford, Parks Road, Oxford, OX1 3PJ, United Kingdom}
\author{Andrew Forbes}
\affiliation{School of Physics, University of the Witwatersrand, Private Bag 3, Wits 2050, South Africa}
\date{\today}
\date{\today}


\begin{abstract}
\noindent \textbf{Skyrmions are topologically protected quasiparticles, originally studied in condensed-matter systems and recently in photonics, with great potential in ultra-high-capacity information storage. Despite the recent attention, most optical solutions require complex and expensive systems yet produce limited topologies. Here we demonstrate an extended family of quasiparticles beyond normal skyrmions, which are controlled in confined photonic gradient-index media, extending to higher-order members such as multiskyrmions and multimerons, with increasingly complex topologies. We introduce new topological numbers to describe these complex photonic quasiparticles and propose how this new zoology of particles could be used in future high-capacity information transfer. Our compact creation system lends integrated and programmable solutions of complex particle textures, with potential impacts on both photonic and condensed-matter systems for revolutionizing topological informatics and logic devices.}
\end{abstract}

\maketitle

\noindent In the current age of information explosion, the pursuit of next-generation information carriers and data storage is endless, eminently benefiting our daily lives. Since Tony Skyrme proposed a model to unify a large class of fundamental particles by methods of topology in the 1960s~\cite{skyrme1962unified}, known today as skyrmions, these have emerged as one of the most potential information carriers due to their topologically robust spin textures localized in ultra-small regions, in particular, which has revolutionized the high-density data storage technologies in magnetic materials in recent years~\cite{bogdanov2020physical,fert2017magnetic,bernevig2022progress,han2022high}. In order to further expand the capacity in informatic applications, it is highly desired to manipulate complex quasiparticles with higher-order topological textures in addition to the fundamental skyrmions~\cite{gobel2021beyond}. For instance, the meron textures with fractional skyrmion numbers can be controlled in chiral magnets~\cite{yu2018transformation}, and recently in antiferromagnets at room temperature~\cite{jani2021antiferromagnetic}. Novel quasiparticles such as skyrmion bags~\cite{foster2019two}, skyrmion bundles~\cite{tang2021magnetic}, and skyrmion braids~\cite{zheng2021magnetic} with large range control of skyrmion numbers were created in magnetic materials. Moreover, exotic spatially knotted quasiparticle of hopfions~\cite{ackerman2017static,liu2020three,liu2022emergent}, torons~\cite{ackerman2017diversity,poy2022interaction}, and heliknotons~\cite{tai2019three}, characterized by complex topologies were designed in magnets, colloids, and chiral liquid crystals. The realization of novel quasiparticles in diversified condensed-matter systems has triggered the development of static or quasi-static ultra-density data storage techniques. However, using topological quasiparticles for long-range secure information transfer remains a considerable challenge.

Optical or photonic skyrmions, which have recently been realized, provide new degrees of freedom in terms of the construction of topological quasiparticles~\cite{shen2022topological}. Following the initial demonstrations of the formation of photonic skyrmions  in  surface  plasma by evanescent electromagnetic fields~\cite{tsesses2018optical,davis2020ultrafast}, as well as optical spins~\cite{du2019deep,dai2020plasmonic}, photonic skyrmions have also been generated in optical polarization and Stokes fields in free space~\cite{shen2021generation,sugic2021particle,shen2023topological}, electromagnetic fields in space-time~\cite{shen2021supertoroidal,zdagkas2021observation}, pseudospins in nonlinear crystals~\cite{karnieli2021emulating}, and so on. However, previous studies have demonstrated a very limited number of topological states, often requiring the use of complex and expensive systems. Moreover, practical information transfer protocols based upon photonic skyrmions are still to be explored and highly desired.

\begin{figure*}[t!]
	\centering
	\includegraphics[width=\linewidth]{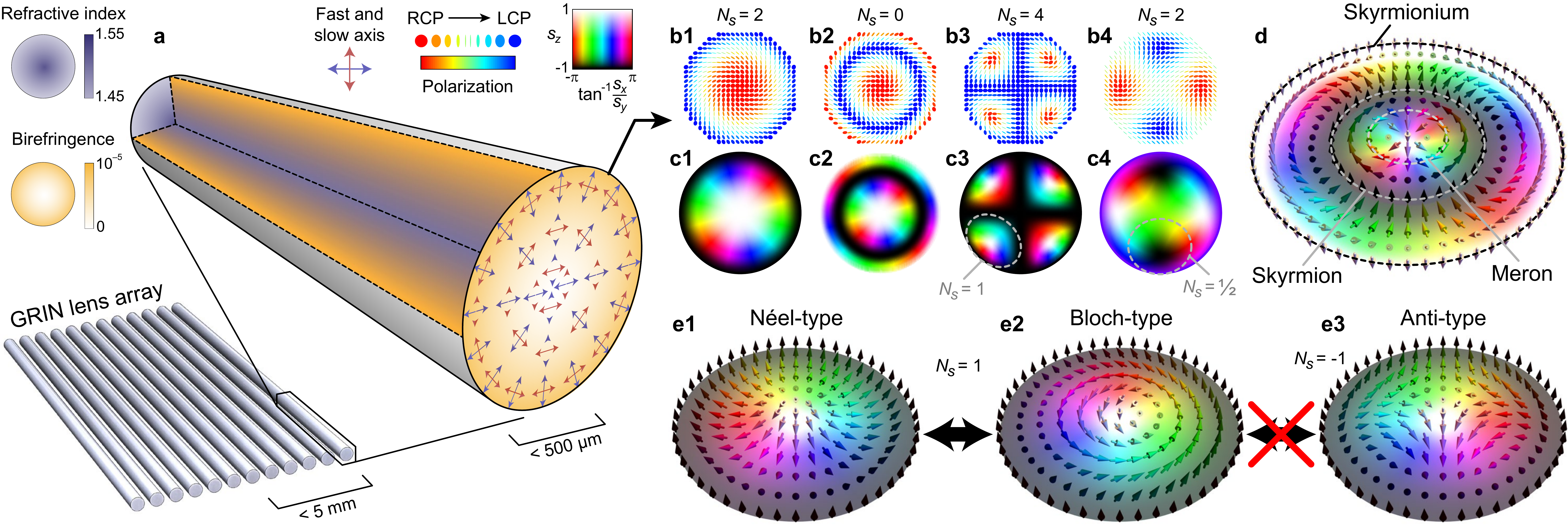}
	\caption{\textbf{GRIN lens arrays for photonic skyrmion generation.} \textbf{a}, Schematic of the GRIN lens and GRIN lens array, with radial and azimuthal distributions of the fast/slow axes. The radius of the cross-section of the lens can be less than 500 microns. \textbf{b,c}, The diversified polarization vectorial light fields corresponding to the different complex photonic quasiparticles generated can be controlled by cascades of GRIN lenses, represented by (\textbf{b}) polarization ellipse distributions and (\textbf{c}) Stokes vector distributions, which include (\textbf{b1,c1}) a skyrmion of $N_s=2$, (\textbf{b1,c1}) a skyrmion of $N_s=0$, (\textbf{b2,c2}) a skyrmionium of $N_s=0$, (\textbf{b3,c3}) a quadruskyrmion of $N_s=4$ comprising four elementary skyrmions of $N_s=1$, and (\textbf{b2,c2}) a quadrumeron of $N_s=2$ comprising four elementary merons of $N_s=1/2$. \textbf{d}, Conceptual schematic of the vector texture of a second-order skyrmionium ($N_r=2$, $N_s=0$), which includes skyrmion ($N_r=1$, $N_s=2$) and meron structure ($N_r=1/2$, $N_s=1$) in its subspace. \textbf{e}, The topological protection of the photonic skyrmions, where (\textbf{e1}) a N\'eel-type skyrmion can be transformed into (\textbf{e2}) a Bloch-type skyrmion upon propagation and resilient to perturbation, under the same skyrmion number of $N_s=1$. It is, however, impossible to be transformed into (\textbf{e3}) an anti-skyrmion of different skyrmion number of $N_s=-1$.} 
	\label{f1}
\end{figure*}

Here, we demonstrate both experimentally and theoretically an extended class of photonic quasiparticles constructed by vectorial structured light in gradient-index (GRIN) lenses~\cite{he2019complex,he2022revealing}. The quasiparticles that are formed using vectorial-structured light exhibit novel forms of higher-order configurations beyond elementary skyrmions with increasingly complex topology and geometry. These complex quasiparticles can be characterized by multiple topological numbers and their diversified topological states can be controlled by cascaded GRIN lenses~\cite{he2019complex}. To illustrate the benefit of this zoology of topological states, we propose a protocol for high-capacity secure encryption, where the multiple topological indices of diversified quasiparticles in a compact GRIN lens array are used to encode information that are robust to environmental perturbations, leading to an alphabet-bit encoded quasiparticle array for high-security optical encryption. In addition, the information capacity in this scheme can be flexibly spanned by arranging the GRIN-lens or quasiparticle array. This work paves the way to the exploitation of topological photonic quasiparticles for next generation information technology, while our novel compact configuration offers the potential to be integrated with on-chip solid state storage devices.

\section{Results}
\noindent
\textbf{Vectorial optics of GRIN lenses.} GRIN lenses are a kind of optical element that possesses spatially varying refractive index profile as well as birefringence, which enable focusing, imaging, and shaping of light through a rod-like structure~\cite{barretto2009vivo,kim2012fabrication,he2019complex,forbes2019common}, as shown in Fig.~\ref{f1}. GRIN lenses can be fabricated with diameters on the sub-millimeter scale, which means that GRIN lens arrays can be formed with dimensions of the order of millimeters (left-bottom insert of Fig.~\ref{f1}). Each GRIN lens can be comprised by different cascaded segments with different refractive index gradient designs, together with segments of quarter wave plates (QWP) and half wave plates (HWP). If we input a plane-wave into one side of a cascaded GRIN lens, the output will be light with spatially varying polarization patterns~\cite{he2019complex,forbes2019common}. Here we will show that through judicious assembly of the cascades of GRIN lenses combined with tuning the polarization of the input light, we can obtain a wide variety of high-order photonic skyrmions with complex topologies.\\[4pt]

\noindent
\textbf{Topological photonic quasiparticles.} A skyrmion refers to a three-component real vector or spin texture mapped from a 2-sphere to a localized 2D real space, donated as $\mathbf{s}(x,y)=[s_x(x,y),s_y(x,y),s_z(x,y)]$, with basic topology characterized by the skyrmion number $N_s$ (which counts how many times the spins can wrap around a 2-sphere with full azimuth):
\begin{equation}
N_s=\frac{1}{4\pi }\iint_\sigma{\mathbf{s}\cdot \left( \frac{\partial \mathbf{s}}{\partial x}\times \frac{\partial \mathbf{s}}{\partial y} \right)}\text{d}x\text{d}y
\end{equation}
where $\sigma$ represents the particle confined region. The photonic skyrmions or quasiparticles can be constructed by the polarization Stokes vectors, i.e. the vector defined by the three Stokes parameters, of the obtained structured light, analogous to the magnetic spin used to construct magnetic skyrmions~\cite{shen2022topological,shen2021topological,gao2020paraxial}. Figure~\ref{f1}\textbf{b1} shows a theoretical result of polarization ellipse distribution of a skyrmionic light field generated by a customized GRIN lens, the stokes vector field of which shows an fundamental skyrmion of $N_s=2$ (Fig.~\ref{f1}\textbf{c1}). Note that, for the fundamental skyrmion, the signal of skyrmion number was decided by its central spin of spin-up ($+$) or spin-down state ($-$), which is defined as the polarity, $N_p=\text{sgn}(N_s)$, and the absolute value of skyrmion number determines the vortex charge of transverse $(x,y)$ component distribution, defined as the vorticity, $N_v=|N_s|$. The result of Fig.~\ref{f1}\textbf{c1} shows and central spin of spin-up and vortex charge of transverse distribution of 2, thus resulting into $N_s=2$. 

In addition to the fundamental skyrmion, cascaded GRIN lens can also generate generalized quasiparticle, for instance, the skyrmionium~\cite{zhang2016control}, a skyrmion radially nested with another one with opposite topological number, resulting into skyrmion number of $N_s=(+2)+(-2)=0$ (Figs.~\ref{f1}\textbf{b2} and \ref{f1}\textbf{c2}). If the radially nested number, $N_r$, is larger than two, the extended configurations are previously called $N_r\pi$-skyrmions or target skyrmions~\cite{song2019field,zheng2017direct}, which, we argue, can also be easily realized in our cascaded GRIN lens system. The versatility of GRIN lenses that are cascaded to form arrays allows us to access novel forms of photonic quasiparticles that are challenging to observe in other systems, for instance, the exotic multiskyrmions and multimerons, comprising multiple elementary skyrmions and merons in subspace with high symmetry. Figures~\ref{f1}\textbf{b3} and \ref{f1}\textbf{c3} demonstrate a quadruskyrmion with four elementary skyrmions of unit skyrmion number, resulting into total skyrmion of $N_s=4$. Figures~\ref{f1}\textbf{b4} and \ref{f1}\textbf{c4}, on the other hand, demonstrate a quadrumeron with four elementary merons of half skyrmion number, resulting into total skyrmion of $N_s=2$. To describe the symmetry in multiskyrmions and multimerons, we define a number of centrality, $N_c$, which counts how many spin-up or -down center points there are. The quadruskyrmion and quadrumeron are both of centrality of $N_c=4$. On the other hand, the skyrmions, skyrmioniums, and target skyrmions are always of centrality of $N_c=1$. Figure~\textbf{d} conceptually demonstrate the elementary relationship among a skyrmionium, a skyrmion, and a meron.

The photonic skyrmions possess the property of topological protection, meaning that the skyrmion number remains invariant upon propagation evolution and is robust to environmental perturbations~\cite{nape2022revealing}. Once a skyrmion is generated by a GRIN lens, its topology will be protected upon further propagation in isotropic media or coupled in free space~\cite{shen2021topological,gao2021erratum}. For instance, an N\'eel-type skyrmion with hedgehog-like texture (Fig.~\ref{f1}\textbf{e1}) can evolve into a Bloch-type skyrmion with vortex-like texture (Fig.~\ref{f1}\textbf{e2}), provided they are of a same skyrmion number $N_s=1$, however, it is impossible for such a skyrmion to be transformed into an antiskyrmion of opposite skyrmion number $N_s=-1$ (Fig.~\ref{f1}\textbf{e3}). The mechanism of topological protection works for arbitrary extended complex quasiparticles.
\\[4pt]

\begin{figure*}[t!]
	\centering
	\includegraphics[width=0.8\linewidth]{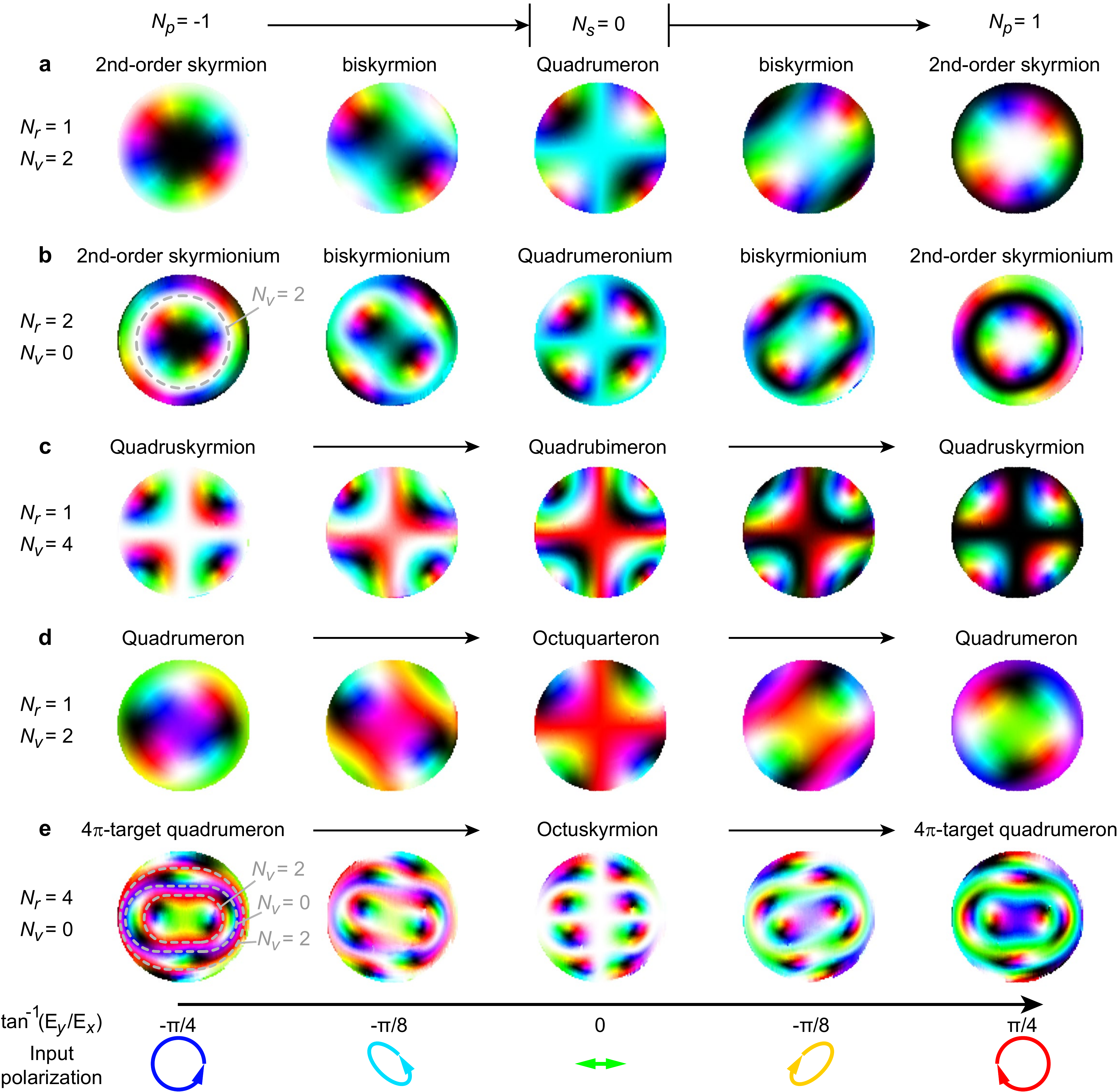}
	\caption{\textbf{Generation and topological transformation of a wide variety of photonic quasiparticles using cascaded GRIN lenses of different configurations} (see also Supplementary Information): \textbf{a}, a skyrmion of $N_r=1$ and $N_v=2$ is transformed into a skyrmion of the same radiality and vorticity but opposite polarity, \textbf{b}, a skyrmionium of $N_r=2$ and $N_v=0$ is transformed into the skyrmionium of the same radiality and vorticity but opposite polarity, \textbf{c}, a quadruskyrmion of $N_r=2$ and $N_v=4$ is transformed into a quadruskyrmion of the same radiality and vorticity but opposite polarity, \textbf{d}, a quadrumeron of $N_r=1$ and $N_v=2$ is transformed into a quadrumeron of the same radiality and vorticity but opposite polarity, and, \textbf{e}, a target quadrumeron of $N_r=4$ and $N_v=0$ is transformed into a target quadrumeron of the same radiality and vorticity but opposite polarity, by tuning the input light polarization from LCP to RCP.} 
	\label{f2}
\end{figure*}
\begin{figure*}[t!]
	\centering
	\includegraphics[width=0.9\linewidth]{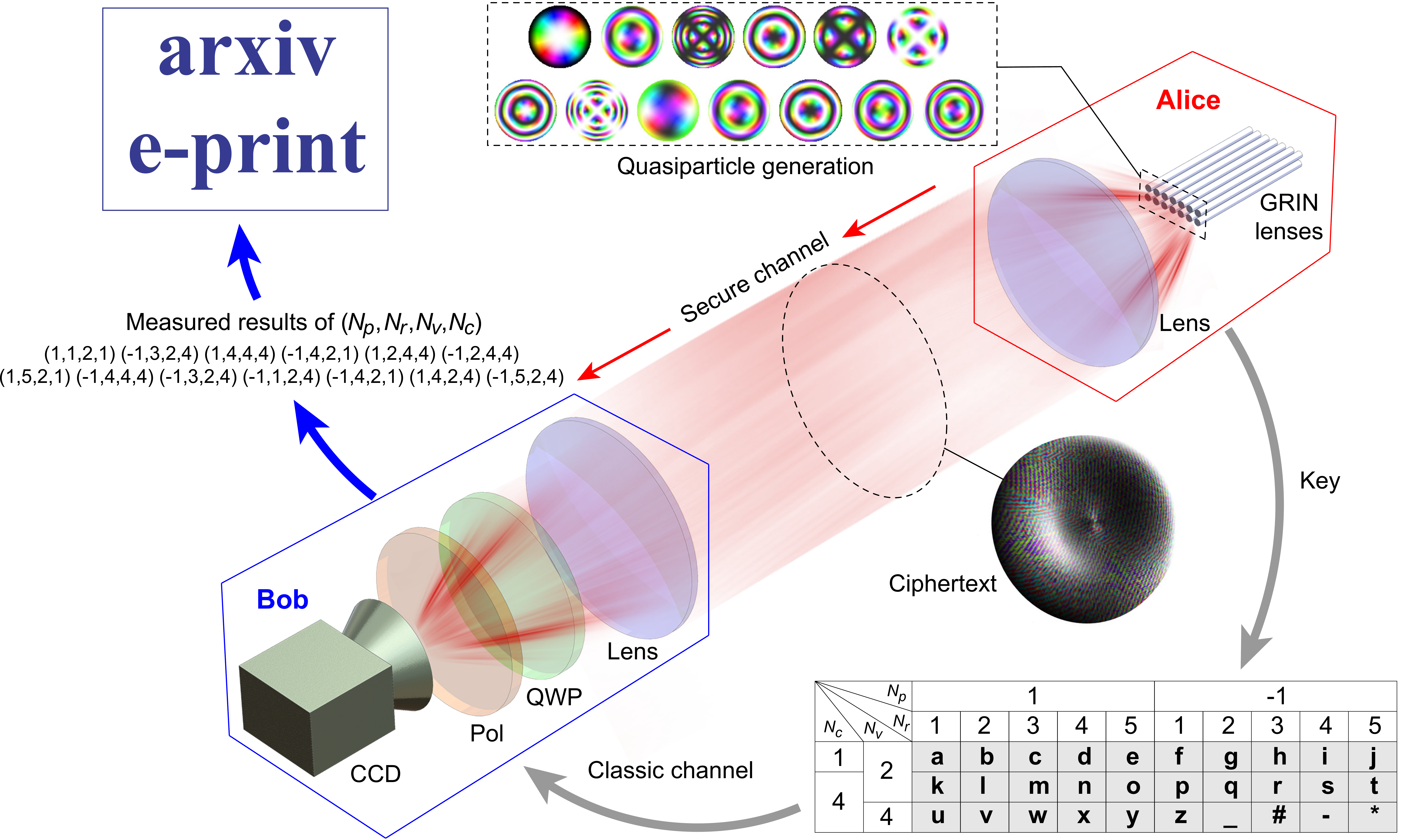}
	\caption{\textbf{Photonic quasiparticle encryption.} Alice holds the cascaded GRIN lens array, which can create high-density arbitrary photonic quasiparticles with on demand polarity, radiality, vorticity, and centrality, in a millimetric small volume. Then Alice relays the photonic quasiparticle array from the Fourier plane to the far-field by a lens, and in the form of s a ciphertext into a free-space secure channel, in which the quasiparticles cannot be detected due to the spatially overlapped superposed state. In parallel, Alice sends a key to Bob via a classic channel which contains the map of the topological numbers in relation to alphanumerical characters. Bob then receives the ciphertext using a Fourier lens to observe the quasiparticle array in the Fourier plane. By apply the detection method as outlined in Methods to measure the four topological numbers of each quasiparticle which are then translated into the plaintext (“arxiv e-print”) using the key only known to Alice and Bob.} 
	\label{f3}
\end{figure*}

\noindent
\textbf{Experimental generation and topological control.} We have designed a series of GRIN lens cascades to experimentally generate a diverse set of complex photonic quasiparticle with controlled multiple topological numbers in addition to skyrmion number $N_s$, which include polarity $N_p$, vorticity $N_c$, radiality $N_r$, and centrality $N_c$. Firstly, we show the topological control of the fundamental skyrmions. Using a GRIN lens segment with input light of left-handed circular polarization (LCP), we can create a 2nd-order skyrmion of $N_p=-1$ and $N_v=2$, see the left of Fig.~\ref{f2}\textbf{a}. To control the polarity $N_p$, we tune the input light polarization from LCP to linear polarization then to right-handed circular polarization (RCP), correspondingly, the skyrmion evolves into biskyrmion and quadrumeron configurations as intermediate states and finally into the skyrmion with opposite polarity ($N_p=1$ and $N_v=2$, the right of Fig.~\ref{f2}\textbf{a}). Note that, for the case of linear polarization input, the output quasiparticle will exhibit a singular symmetry inducing skyrmion number of zero, $N_s=0$, where the polarity, $N_p=\text{sgn}(N_s)$, cannot be defined.

To control radiality, we apply a cascade of two identical GRIN lens segments, which results into a skyrmionium, see the left of Fig.~\ref{f2}\textbf{b}. Polarity control by tuning input polarization works for this case as well, and the corresponding results are similar to that presented in Fig.~\ref{f2}\textbf{a} but with a layer of radially nested structure, resulting into a skyrmion number of zero. 

To obtain more complex quasiparticle with increased skyrmion number, we show experimental results of quadruskyrmion by applying a cascaded configuration of ``GRIN+HWP+GRIN'', see the left of Fig.~\ref{f2}\textbf{c}, which comprises four skyrmions with same unit vorticity. Polarity control can also be applied to the quadruskyrmion. In addition, by switching to the cascade of ``QWP+HWP+GRIN'', the quadrumeron can be generated comprising four elementary merons of the same half charge, see Fig.~\ref{f2}\textbf{d} (distinct from the quadrumeron in the middle of Fig.~\ref{f2}\textbf{a}).

By varying the design of the cascaded GRIN lens array, increasingly complex topologies can be accessed. When we cascade multiple times of the style of ``QWP+HWP+GRIN'', we can generate quasiparticles of multi-radially-nested quadrumeronium and target quadrumeron. Figure~\ref{f2}\textbf{e} shows an experimental quadrumeronium and target quadrumeron with radiality of $N_r=4$ by a multiple of four cascades of the configuration described previously, i.e. a four-times nested matryoshka-like structure based on a basic central structure of quadrumeron as in Figure~\ref{f2}\textbf{d}. 
Based on the principle outlined above, the quasiparticles that can be accessed are far greater than those demonstrated here. For example, we could easily generate and control the following photonic skyrmions: quadruskyrmionium, octuskyrmion, octumeron, octuskyrmionium, octumeronium, etc. Furthermore, there is still a huge space to explore in terms of the creation of more exotic and complex quasiparticles that have not existed before.
\\[4pt]

\noindent
\textbf{Topological optical encryption.} Based on the complex photonic quasiparticles characterized by multiple topological numbers and confined in very small volume, we conceptionally demonstrate a new high-density long-range information transfer protocol that exploits these quasiparticles as information carriers, as shown in Fig.~\ref{f3}, Alice holds a GRIN lens array system, which can generate different photonic quasiparticles with customized topological numbers in terms of the polarity $N_p$, vorticity $N_c$, radiality $N_r$, and centrality $N_c$, as demonstrated in last section. Alice is able to prepare a secret key, i.e., a map from quasiparticle topological numbers to plaintext. Due to the advantage of the vast range of controllable topological numbers of the quasiparticle, the plaintext can be encoded by a large number of bits, for instance, the key for 30-bit encoding is shown in the bottom right of Fig.~\ref{f3}, which covers a full alphabet. All the quasiparticles corresponding to the bits can be theoretically generated in our cascaded GRIN lens arrays, see Methods. Note that this scheme can be easily extended into higher-dimensional bit encoding by exploiting the control of higher topological numbers of multiskyrmions and the information bandwidth or capacity (speed) in the channel can also be extended by simply increasing the GRIN lenses.

An example of the preparation of the photonic quasiparticles for the plaintext message ``arxiv e-print'' that Alice would like to send is shown in the top-middle of Fig.~\ref{f3}, which corresponds to the customized quasiparticles with different combinations of topological numbers $(N_p,N_r,N_v,N_c)$ of $(1,1,2,1)$, $(-1,3,2,4)$, $(1,4,4,4)$, $(-1,4,2,1)$, $(1,2,4,4)$, $(-1,2,4,4)$, $(1,5,2,1)$, $(-1,4,4,4)$, $(-1,3,2,4)$, $(-1,1,2,4)$, $(-1,4,2,1)$, $(1,4,2,4)$, $(-1,5,2,4)$. Next, Alice generates the corresponding quasiparticles array, and couples them into free space as a secure channel by a Fourier lens and sends to Bob (see Method). In the secure channel, the light behaves as a ciphertext, i.e. a spatially superposed state of various quasiparticles due to Fourier lens transformation, where it is impossible to detect information by a third party without a specific polarimetry device, so as to ensure security. At the receiver side, Bob need to prepare another lens to focus the ciphertext light onto the Fourier plane, where the quasiparticle array is imaged and a CCD camera is used to capture the image. Between the CCD and lens is a Stokes polarimetry system comprising a QWP and a polarizer, which is used to measure distributions of all components of Stokes vector, see details in Methods. After reconstruction of quasiparticle array, Bob can read the combinations of the four topological numbers $(N_p,N_r,N_v,N_c)$, assisted by simple calculations, see details in Methods. Finally, Alice need to send the key to Bob by another classic channel, where a third party also cannot get any readable information. Once the key has been received, Bob can translate the measured results for the quasiparticle topological numbers into readable plaintext, and this massage is only known between Alice and Bob. In this example, we have used ``arxiv e-print'' as the plaintext just for simplification, which can be replaced with any arbitrary length texts, exploiting the diversity of quasiparticles and their range of controlled topological numbers.

\section{Discussion}
Here we have created a novel family of topological quasi-particles controlled by photonic GRIN lens cascades, extending the fundamental skyrmions to exotic radially nested skyrmioniums, multiskyrmions and multimerons, which possess sophisticated spin textures  characterized by multiple topological numbers. The scheme presented here could be further extended in terms of diversity of the photonic quasiparticles by taking advantage of the plethora of potential designs of the cascades of the GRIN lenses. In our experiment, we repeatedly cascade the same combination of segments to control the radially nested structure, but in the future, it would also be interesting to explore alternative hybrid cascades of different combination segments and to identify a mechanism to control more complex forms of quasiparticles. 

The versatility of cascaded GRIN lens design will open new technological directions to continually explore and discover new kinds of topological quasiparticles with sophisticated textures and properties. In addition to the proposed concept of encryption, the quasiparticles still have great potential for other informatic applications such as optical ultra-capacity communications~\cite{wan2022divergence,pryamikov2022rising,willner2021orbital}, high-dimensional multi-party secret sharing~\cite{pinnell2020experimental}, and ultra-sensitive ultraprecise sensing and metrology~\cite{li2022highly,yuan2019detecting}, and more. Here, the topological photonic quasiparticles always have the advantage of much higher information or data density in contrast to other carries, due to their multiple controlled degrees of freedom in particle-like unit volumes. For instance, using orbital angular momentum (OAM) of light as a carrier can also achieve same encryption, as OAM can also possess high-dimensional topological numbers with chirality control, equivalent to the vorticity with polarity control of skyrmions. While, the multiskyrmions open additional topological numbers such as centrality and radiality and controlled in more condensed spatial volume, promising step towards future ultra-capacity optical communications and informatics. In addition, in contrast to the scalar OAM light, the vectorial topologies of photonic multi-skyrmions can provide strong resilience against the perturbations in complex media such as turbulence~\cite{nape2022revealing}, promising the improvement of robustness in communication networks.

Finally, the new photonics quasiparticles proposed and discovered here in photonic fields are also highly desired for other communities and in other condensed-matter systems. On the one hand, quasiparticles were transferred to photonics from magnetic and condensed-matter systems to created interdisciplinary studies, on the other hand, photonic methods can inspire new forms of quasiparticles not existed before in any system, leading to new fundamental physical effects that could be explored and new applications such as topological optoelectronic devices. 

\section{Methods}
\noindent
\textbf{Stokes polarimetry of structured light.} The experimental reconstruction of the optical quasiparticles is achieved through Stokes polarimetry, more specifically, through the reconstruction of the distributions of three normalized Stokes parameters $s_1$, $s_2$ and $s_3$, which were computed from a set of polarization projection and intensity measurements. Figure~\ref{fm} demonstrates the setup for Stokes vector field measurements. When the generated Skyrmionic beams from a GRIN lens or its cascades passing through a quarter waveplate (QWP) and a polarizer (P), spatial variance intensity pattens are recorded by a CCD camera. We used the QWP, P and the CCD to form a Stokes polarimeter to measure the polarization distribution by rotating the QWP to four different angles. This is a well-known process that can be found in previous technical reviews~\cite{he2021polarisation,he2022revealing}. The principal equations for calculation of the polarization field are given by:
\begin{equation}
S^n_\text{out}=M_\text{P}\cdot M^n_\text{QWP}\cdot S_\text{in}=M_\text{P}\cdot M^n_\text{QWP}\cdot(A^{-1}\cdot I)
\end{equation}
where $S_\text{in}$ is the Stokes vector of the incident light field, $M_\text{P}$ and $M^n_\text{QWP}$ are Mueller matrices of the corresponding P and QWP, $M^n_\text{out}$ is the output Stokes vector for the $n$-th fast axis orientation state of the QWP. $M^n_\text{QWP}$ is the Mueller matrix of the QWP for the $n$-th fast axis orientation, $A$ is an instrument matrix, which is derived from $M_\text{P}M^n_\text{QWP}$. $I=A\cdot S_\text{in}$ is the intensity information recorded by the CCD camera. Note that for a 2D distributed Stokes vector field $S_\text{in}(x,y)$, it consists of $[1,s_1(x,y),s_2(x,y),s_3(x,y)]$, where $s_1$, $s_2$, and $s_3$ are the vector components of Stokes vector~\cite{he2022revealing}.
\\[8pt]
\begin{figure}
	\centering
	\includegraphics[width=0.8\linewidth]{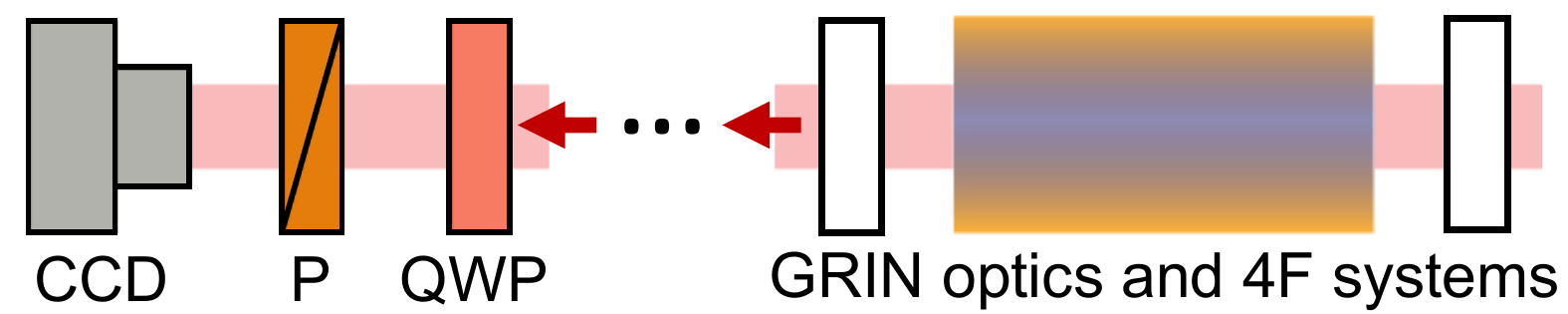}
	\caption{\textbf{Schematic setup of the configuration used for Stokes polarimetry.} P: polarizer; QWP: quarter waveplate; CCD: detector. The remaining components relate to the GRIN lens (or cascades of GRIN lenses). The cascades are combined with different types of GRIN lenses and interstitial optical elements (such as a 4f system). Note that although the GRIN lens itself can be an imaging device, we use a 4f system for extended imaging purposes.} 
	\label{fm}
\end{figure}
\begin{table*}[!htbp]
\caption{Cascade design for secret key preparation}
    \label{tab}
    \centering
    \begin{tabular}{|c|c|c|c|c|c|}
    \hline
    \diagbox{$(N_c,N_v)$}{$N_r$} & 1 & 2 & 3 & 4 & 5 \\
    \hline
    $(1,2)$  & GRIN & GRIN+GRIN & GRIN+GRIN+GRIN & GRIN$\times$4 & GRIN$\times$5 \\
    \hline
    $(4,2)$  & C1=QWP+HWP+GRIN & C1+C1 & C1+C1+C1 & C1$\times$4 & C1$\times$5 \\
    \hline 
    $(4,4)$  & C2=GRIN+HWP+GRIN & C2+C2 & C2+C2+C2 & C2$\times$4 & C2$\times$5 \\
    \hline 
    \end{tabular}
\end{table*}
\begin{figure*}[t!]
	\centering
	\includegraphics[width=\linewidth]{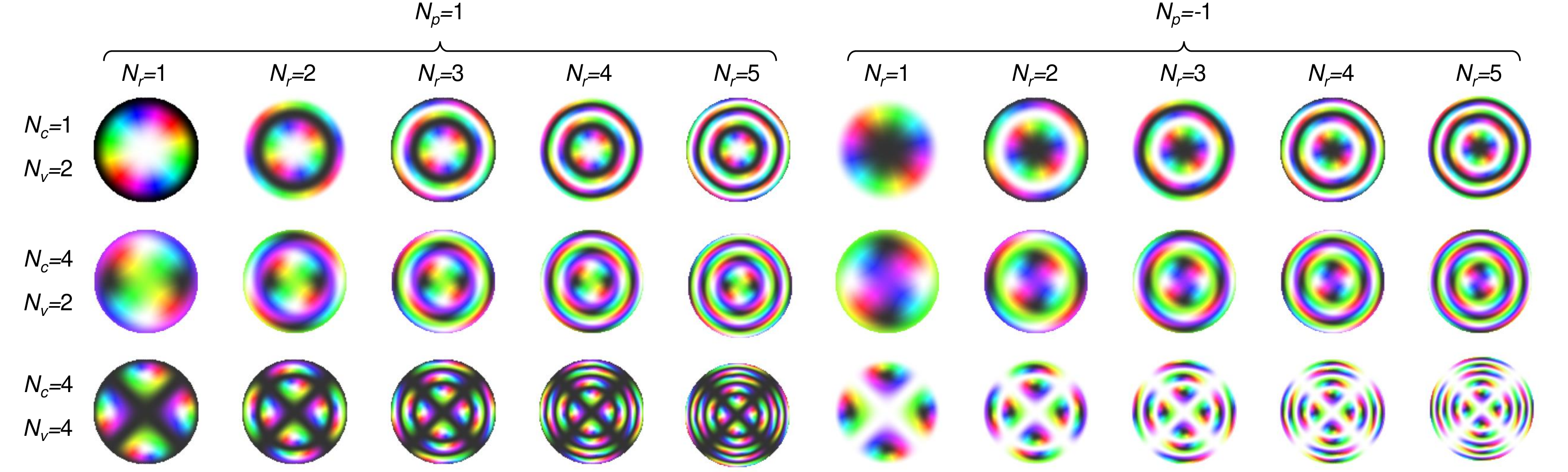}
	\caption{The photonic quasiparticles of controlled topological numbers $(N_p,N_r,N_v,N_c)$ as encoding bits for secret key preparation. The left and right groups represent the quasiparticles with opposite polarity $N_p=\pm 1$. Columns in each group represent the control of different values of radiality from $N_r=1$ to $5$. Three rows are for three different controls of centrality and vorticity, $(N_c,N_v)=(1,2)$, $(4,2)$, and $(4,4)$.} 
	\label{fm}
\end{figure*}

\noindent
\textbf{Detection of quasiparticle topological numbers.} After reconstruction of the Stokes field of a quasiparticle,  $\mathbf{s}(x,y)=[s_1(x,y),s_2(x,y),s_3(x,y)]$, the number of $N_r$ can be easily decided by counting the radially nested structures in which. For the cases of $N_r>1$, we always care about the basic structure in the center, we refer to here as the core structure, and the outer layers are just the same as the core structure only with staggered polarity. Then, we care about the number of $N_c$ of the core structure, i.e. How many spin-up or -down center points in the core region exist, which is equivalent to how many singularities of the transverse component $[s_1(x,y),s_2(x,y)]$ are contained in the core region. This singularity counting problem can be solved by an algorithm for singularity searching of a complex multi-singularity light field~\cite{wang2022deep}. To further characterize structural details, the skyrmion density distribution of the quasiparticle is calculated by $\rho_s(x,y)=\mathbf{s}(x,y)\cdot[\partial_x\mathbf{s}(x,y)\times\partial_y\mathbf{s}(x,y)]$. If $N_r>1$, the skyrmion density distribution will also correspondingly show radially nested structure. Next, we 
integrate the skyrmion density to the core region and the skyrmion number $N_s$ of the core structure is obtained. Finally, the polarity and vorticity can be obtained by $N_p=\text{sgn}(N_s)$ and $N_p=|N_s|$, and all the four numbers $(N_p,N_r,N_v,N_c)$ required to characterize the photonic quasiparticles (multiskyrmions and multimerons) are measured, see detailed analysis and calculation results of experiments in Supplementary Information.
\\[8pt]
\noindent
\textbf{Secret key preparation.} To achieve the functionality of the key shown in Fig.~\ref{f3}, we need to build the channel to controllably emit corresponding photonic quasiparticle. In this channel, the sender hold multiple segments of HWP, QWP, and GRIN lenses. The cascade designs corresponding to the required topological numbers of $(N_r,N_v,N_c)$ are shown in Table~\ref{tab}. The polarity $N_p=\pm 1$ is controlled by the input light with polarization of either RCP or LCP. The corresponding quasiparticles are shown in Fig.~\ref{fm}. 
\\[8pt]

\bibliographystyle{naturemag}

\noindent
\textbf{Acknowledgments} \\ C.H. would like to thank the support of the John Fell Fund from University of Oxford, and the Junior Research Fellowship from St John’s College, University of Oxford.
\\[8pt]
\noindent
\textbf{Author contributions}\\
Y.S. conceived the idea and wrote the paper, C.H. designed the GRIN lens cascades, performed simulation and experiment of the GRIN lens based skyrmionic beam generation, with the input of Z.S., B.C., H.H., Y.M., J.A.J.F., S.J.E., S.M.M., and M.J.B. together. Y.S, C.H., S.M.M., and A. F. contributed to data analysis and manuscript revision.
\\[8pt]
\noindent
\textbf{Competing interests}\\
The authors declare no competing interests.

\bibliography{sample}

\end{document}